\documentclass[a4paper]{article}
\pdfoutput=1
\usepackage{INTERSPEECH2022}
\usepackage{amsmath}

\title{PRISM: Pre-trained Indeterminate Speaker Representation Model for Speaker Diarization and Speaker Verification}
\name{Siqi Zheng, Hongbin Suo, Qian Chen}
\address{
  Speech Lab, Alibaba Group}
\email{\small\texttt{\{zsq174630, tanqing.cq\}@alibaba-inc.com}}

\begin{document}

\maketitle
\begin{abstract}
Speaker embedding has been a fundamental feature for speaker-related tasks such as verification, clustering, and diarization. Traditionally, speaker embeddings are represented as fixed vectors in high-dimensional space. This could lead to biased estimations, especially when handling shorter utterances. In this paper we propose to represent a speaker utterance as ``floating" vector whose state is indeterminate without knowing the context. The state of a speaker representation is jointly determined by itself, other speech from the same speaker, as well as other speakers it is being compared to. The content of the speech also contributes to determining the final state of a speaker representation. We pre-train an indeterminate speaker representation model that estimates the state of an utterance based on the context. The pre-trained model can be fine-tuned for downstream tasks such as speaker verification, speaker clustering, and speaker diarization. Substantial improvements are observed across all downstream tasks.
\end{abstract}

\section{Introduction}
\label{sec:intro}

An embedding is a learned representation of objects in a fixed-dimensional space for the convenience of further analysis. It is a commonly-used concept in the fields of computer vision (e.g. face embedding), natural language processing (e.g. word embedding), and speech processing (e.g. speaker embedding) \cite{DBLP:conf/cvpr/SchroffKP15}\cite{DBLP:journals/corr/abs-1301-3781}\cite{DBLP:conf/interspeech/SnyderGPK17}. Traditionally, an embedding has been established as a fixed vector in the learned latent space. This may lead to biased estimation of embedding. For example, a speaker embedding is extracted from a segment of speech signals. When the segment is long enough, the speaker embedding provides a reasonable approximation of the speaker's voice. When the utterance is short, however, the estimation is highly biased by the content of the speech, as well as other distracting factors.

Therefore, we propose to view representations of speakers in a different way. Instead of establishing speaker embeddings as fixed vectors, we treat them as indeterminate on the set of all possible measurable outcomes. The final embedding state is uniquely determined by the context in which the speaker utterance is evaluated. The context can include, but not limited to, the speech content and acoustic signals of the utterance of interest and those of other utterances it is being compared to. 

In speaker verification, fixed-embedding approaches such as i-vectors, x-vectors, and d-vectors have been proven to perform reasonably well on long utterances since we can collect enough information to estimate a proper representation\cite{DBLP:conf/interspeech/SnyderGPK17}\cite{DBLP:conf/icassp/LeiSFM14}\cite{DBLP:conf/icassp/WanWPL18}. On text-independent short utterances, however, the performance of all above approaches decreased drastically. This is mainly because the speaker embeddings extracted from shorter utterances are biased by non-speaker-related information such as the content of the speech.

Phonetic bottleneck features have been previously used in speaker verification, mostly in a non-contrastive manner\cite{DBLP:conf/icassp/McLarenFL16}\cite{DBLP:conf/odyssey/Lozano-DiezSMGP15}\cite{DBLP:conf/interspeech/RahmanHMFS18}. Hence the phonetic similarity between any two utterances are not directly compared. In \cite{DBLP:conf/interspeech/ZhengS20} the authors use a contrastive approach to regularize the content of the speech by forcing the network to learn whether the closer distances between two short utterances are caused by the similar phonetic contents or the actual vocal features. In this study we generalize the idea to the pre-training of indeterminate speaker representation. The latent speaker representation of each frame is jointly estimated by itself, other frames in the utterance, as well as frames from other utterances in the context. The \textbf{Pr}e-trained \textbf{I}ndeterminate \textbf{S}peaker representation \textbf{M}odel (PRISM) can be fine-tuned on multiple downstream tasks, including speaker verification, speaker clustering, and speaker diarization.

From the speaker diarization point of view, the hidden speaker-related feature of each frame is determined by the information from entire sequence of input, which may comprise arbitrary number of speakers. The mutual inter-dependency is modeled by transformer architecture\cite{DBLP:conf/nips/VaswaniSPUJGKP17}. 

A popular approach of speaker diarization is to run clustering on speaker embeddings extracted from segments of speech \cite{DBLP:conf/icassp/Garcia-RomeroSS17}\cite{DBLP:conf/icassp/LandiniWDBMZMSP20}\cite{DBLP:conf/icassp/SellG15}\cite{DBLP:conf/icassp/ZhengHWSFY21}. In related prior studies, embeddings are treated as fixed vectors and clustering methods such as k-means and agglomerative hierarchical clustering are implemented \cite{DBLP:conf/icassp/Garcia-RomeroSS17}\cite{DBLP:conf/icassp/LandiniWDBMZMSP20}. 

In this work we no longer run clustering on fixed points, since the speaker embeddings are indeterminate. Instead, the relationship between segments are represented by a graph whose edge values are the pairwise similarity between segments. Hence the Laplacian graph can be constructed and graph-based clustering approaches are adopted.

The end-to-end speaker diarization approach is attracting increasing attention \cite{DBLP:conf/asru/FujitaKHXNW19}\cite{DBLP:conf/interspeech/FujitaKHNW19}\cite{DBLP:conf/interspeech/HoriguchiF0XN20}. It trains the network using permutation invariant loss and directly outputs the speaker labels. We show that end-to-end speaker diarization could also benefit from pretraining, as the training objective is fairly similar. In both PRISM and diarization fine-tuning, the goal is to differentiate speaker identities of each frame from a continuous sequential audio input. 

End-to-end speaker diarization approaches have two limitations. First, its performance drops significantly as the number of speakers increases. Second, it is computationally expensive, which makes it unsuitable for long meetings. In this paper we present a speaker diarization system based on PRISM that works well in meetings that last for hours and contain more than 8 speakers. 

Unsupervised learning of speaker verification has long been an important area of research. There are a number of studies trying to leverage the benefits of large amounts of unlabelled speech data to improve speaker verification performance \cite{DBLP:conf/interspeech/WangK17}\cite{DBLP:conf/interspeech/KhanH20}\cite{DBLP:conf/interspeech/ZhengS19}\cite{DBLP:conf/icassp/CaiWL21}. In this work we show that PRISM could be used for unsupervised pre-training and benefit from utilizing unlabeled data.

The rest of the paper is organized as follows. In section 2 we describe the mechanism of PRISM and how it helps to improve downstream tasks such as speaker verification and diarization. Section 3 presents experiments on two corpus, the publicly available NIST SRE corpus and a private corpus to illustrate its effects on unsupervised pre-training. The results and ablation studies are discussed in section 4.

\section{PRISM}

\subsection{Motivation}

Figure 1 illustrates one of the motivations of proposing indeterminate speaker embeddings. Figure 1 (i) visualizes the conventional approach of treating all embeddings as fixed vectors. Embeddings A, B, and C are from different speakers. Embeddings A and A$'$ belong to the same speaker. In most speaker verification task, a threshold is pre-determined and embeddings lying outside the threshold are considered belonging to different speakers. In Figure 1(i) A and A$'$ are misidentified as different speakers while A and B are mistakenly treated as same speakers. This happens more often for shorter utterances as variations are larger. Our goal is to model the embedding to be indeterminate and dependent on the context, as illustrated Figure 1(ii), (iii), and (iv). When being compared to utterance B, the embedding of A$'$ should ideally be set at the proximity as shown in Figure 1(iii). We expect the estimation of A$'$ takes into account the phonetic contents of A$'$ and B so that an unbiased distance between the two utterances can be found. If the same threshold is used, utterance B is now correctly identified as a different speaker. On the other hand, when compared to utterance A, a different state of A$'$ is estimated, which now returns the correct decision, as shown in Figure 1(iv). 

\begin{figure}[h!] 
\centering
  \includegraphics[width=\linewidth]{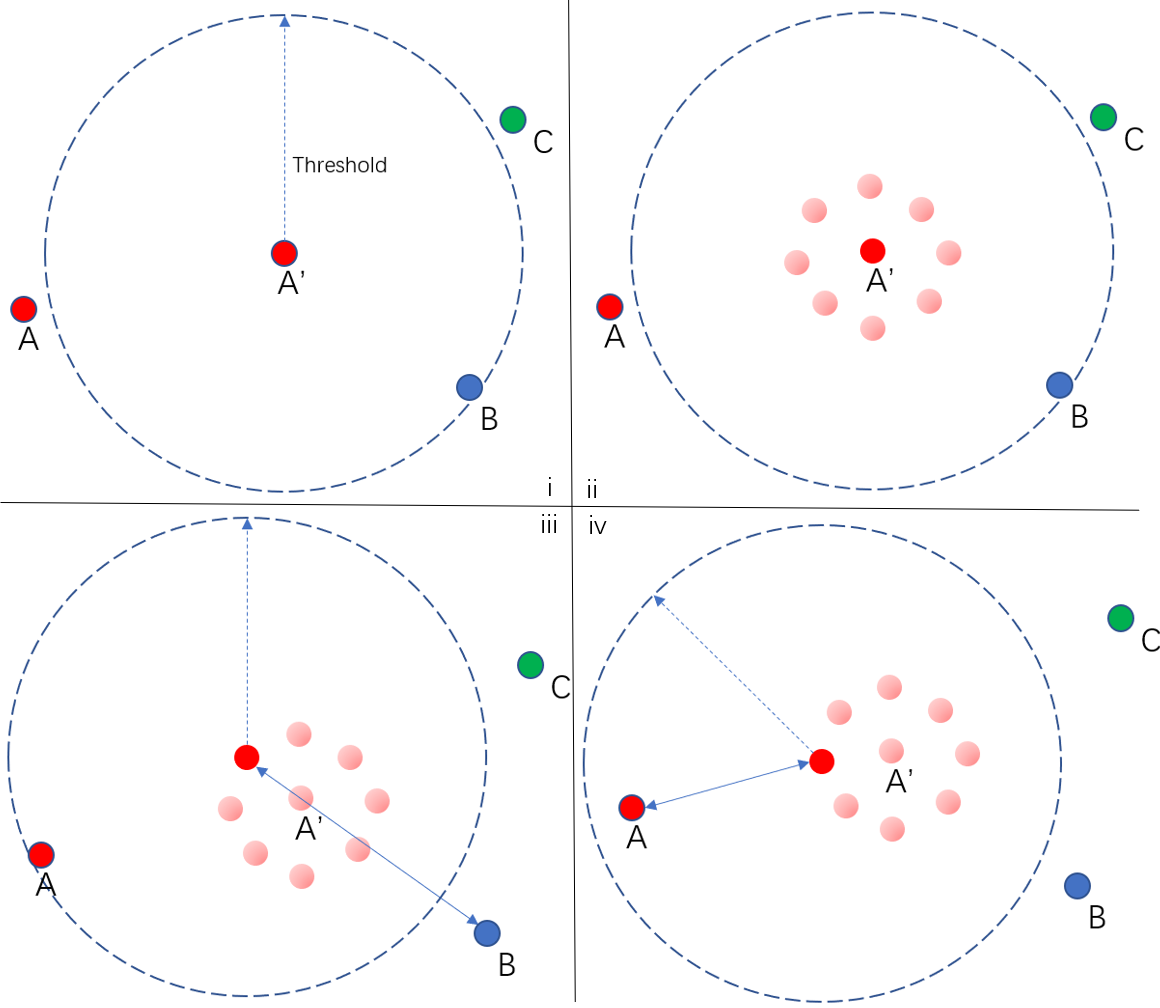}
  \caption{(i) embedding A$'$ is viewed as a fixed vector; (ii) embedding A$'$ is viewed as indeterminate distribution; (iii) the state of A$'$ is determined when compared to utterance B; (iv) a different state of A$'$ is determined when compared to utterance A.}
  \label{motiv}
\end{figure}

\subsection{Pre-training}

\begin{figure*}[t]
  \centering
  \includegraphics[width=\linewidth,height=200pt]{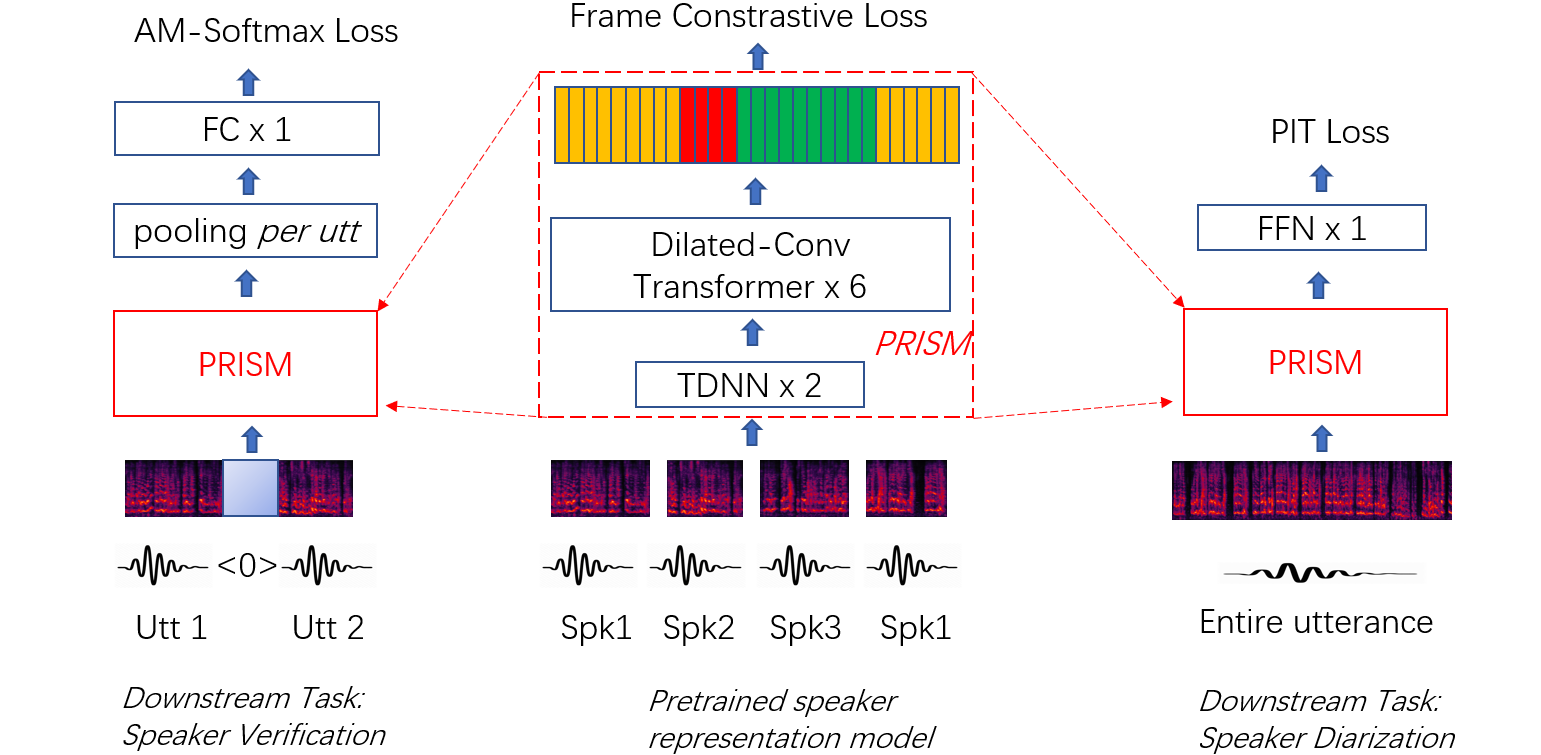}
  \caption{The architecture of PRISM. Two downstream tasks are displayed - speaker verification and speaker diarization. }
  \label{fig:prism}
\end{figure*}

As illustrated in Figure \ref{fig:prism}, PRISM consists of two layers of TDNN and 6 dilated-convolutional transformer encoder layers. Let $T$ be the length of inputs, then the length of outputs after two TDNN layers is $T/4$. This reduces the computational cost of transformer. A dilated convolution layer is added on top of each transformer encoder layer described in \cite{DBLP:conf/nips/VaswaniSPUJGKP17}. This helps to capture local information, which is useful for extracting robust speaker representation. Dilation is used to increase receptive field. For each frame, a receptive field of $\pm 1.4$ seconds is covered to estimate the state of that frame. 

For each frame in the inputs, an ASR bottleneck feature is extracted to represent the phone information using similar approach as \cite{DBLP:conf/interspeech/ZhengS20}. The phonetic feature is concatenated to the acoustic feature. 

Let $\{y_1, \ldots, y_T\}$ be the frame-level outputs from the transformer. We denote $y_{i,u}^{(n)}$ as the output at the $i$th frame belonging to speaker $u$ from the $n$th sample in a batch. Hence we write the pre-training frame-contrastive loss function as 
    \begin{multline}
    \label{eq1}
    \mathcal{L}_{PRISM} =  \frac{1}{N}\sum_{n=1}^N\Bigg(\max \bigg(\frac{1}{P}\sum_{\substack{1\leq i,j\leq T, \\ u\not = v}} y_{i,u}^{(n)}y_{j,v}^{(n)} - \\ \frac{\beta}{Q}\sum_{1\leq i,j\leq T} y_{i,u}^{(n)}y_{j,u}^{(n)} + \alpha, 0\bigg)\Bigg)
    \end{multline}
\noindent where $P$ and $Q$ are the  the number of negative and positive frame-pairs in a sample. $\alpha$ is the margin and $\beta$ controls the relative importance between positive and negative incidence.

Minimizing $\mathcal{L}_{PRISM}$ ensures that frames from the same speakers are close to each other and frames from different speakers are far from each other. It also serves the purpose of regularizing phonetic contents. With explicit introduction of phonetic features, PRISM learns to separate contributions of voice-related information and content-related information.

One underlying assumption of this pre-training design is that any two random utterances drawing from a large enough pool of training data belong to different speakers, and each utterance contains one single speaker. According to our industrial experience, such data are relatively simple and affordable to obtain. This way we assign pseudo-labels to all unlabeled data and make ``unsupervised" pre-training supervised.

\subsection{Downstream task: Verification}

PRISM can be applied to 1-1 speaker verification task on any two utterances $u_1$ and $u_2$. When fine-tuning on 1-1 verification task, the two selected utterances are concatenated with enough zero padding in between to ensure that the receptive field of dilated convolution does not cover each other, as shown in the left branch of Figure 1. Pooling is carried out on each of the two utterances. Two AM-Softmax losses are estimated, each for one of the utterances. The total loss for fine-tuning is
\begin{equation}
\mathcal{L}_{ver} = \mathcal{L}_{AMS1} + \mathcal{L}_{AMS2} + \lambda \mathcal{L}_{PRISM}.
\end{equation}
\noindent $\lambda$ balances the weight between frame-level contrastive loss and utterance level AM-Softmax loss. By minimizing $\mathcal{L}_{ver}$ we aim to satisfy two objectives. On one hand, speaker representation from each frame should take into account all the context information, including acoustic and phonetic information from both utterances. On the other hand, utterance level information is leveraged by AM-softmax to avoid excessive concentration on detailed regions. 

\subsection{Downstream task: Diarization}

The inputs for training PRISM are sequences of utterances from multiple speakers, which makes diarization a naturally suitable downstream application. The objective of diarization fine-tuning is to minimize
\begin{equation}
\mathcal{L}_{diar} = \mathcal{L}_{PIT} + \gamma \mathcal{L}_{PRISM},
\end{equation}
\noindent where $\mathcal{L}_{PIT}$ is the permutation invariant loss used in end-to-end speaker diarization\cite{DBLP:conf/interspeech/FujitaKHNW19}.

\subsection{Speaker diarization system based on PRISM}

\begin{figure}[h!] 
\centering
  \includegraphics[width=\linewidth]{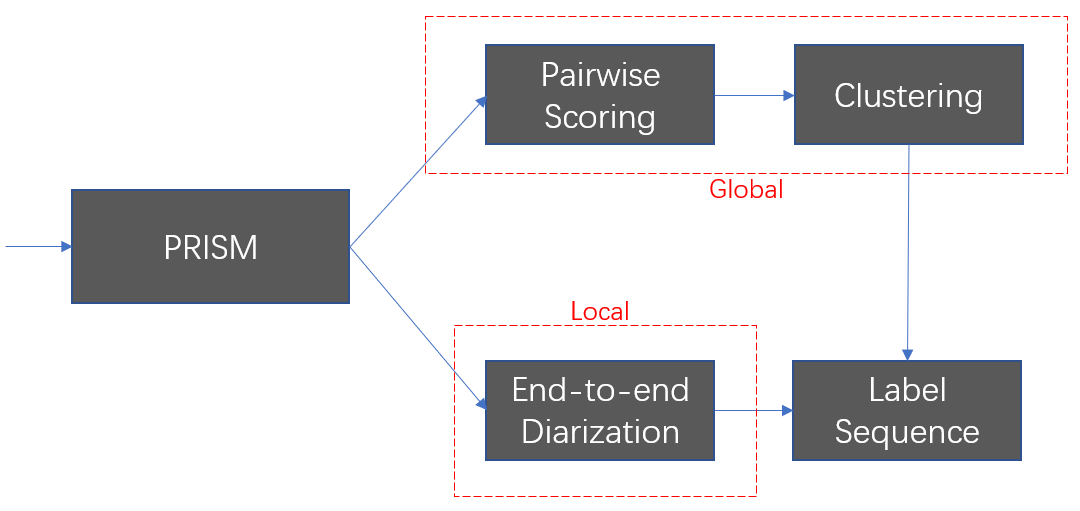}
  \caption{PRISM-based speaker diarization system.}
  \label{flowchart}
\end{figure}

\textbf{A hybrid diarization system.} The clustering-based diarization methods tend to perform better on long input sequences with large number of speakers than end-to-end approaches. We propose to take advantage of each side, by extracting global information using clustering-based methods and handling local information using end-to-end mechanism. Global information includes number of speakers and the clustering results of each segments. Local information includes treatment of overlapping speech, location of exact speaker change point and frame-level speaker labels. All of the above tasks can benefit from PRISM, as illustrated in Figure \ref{flowchart}. 

\textbf{An effective density-based clustering method.} Similar to the conventional clustering-based approach, the inputs are split into short segments. Pairwise similarities are estimated among all segments using PRISM-verification network as described in 2.2. Centroid-based clustering methods such as k-means are no longer appropriate since there is no explicit representation for any single point. Instead, we represent each segment as a node in graph and the edge values between two nodes represent the similarities between two corresponding segments. Hence we formalize clustering problem as finding dense regions separated by sparse regions. We find that density-based approaches such as HDBSCAN\cite{DBLP:journals/jossw/McInnesHA17} work well on speaker clustering task. 

Specifically, we modify the HDBSCAN by estimating density at the proximity of embedding $A$ using PRISM-pairwise similarities, denoted as $\mathcal{D}(\delta_{PRISM}(A),k)$, where $k$ represents the $k$-th nearest neighbor of $A$. The density distribution is then projected from Euclidean space to $\lambda$-space by replacing the distance between $A$ and $B$ by the maximum of the following: 1. $\mathcal{D}(\delta_{PRISM}(A),k)$, 2. $\mathcal{D}(\delta_{PRISM}(B),k)$, 3. $\delta_{PRISM}(A,B)$. This helps to spread points in sparse region to avoid small undesired clumps. 

\textbf{End-to-end post-processing.} The results of clustering reveal some global information of the meeting, such as the number of participants in the meeting, and the assignment of each segments. To handle local refinement such as smoothing and overlapping segments, we use NN-based post-processing approaches as described in \cite{DBLP:conf/icassp/ZhengFY22}.

\section{Experiments}

\subsection{Corpus}

The experiments are conducted on two separate datasets: the publicly-available NIST SRE corpus and a private dataset with more speakers. The NIST SRE training corpus consists of 57,517 utterances from 5,767 speakers in NIST SRE 04-10 corpus. The performance is evaluated on SRE10 Evaluation set. When evaluating the performance on speaker verification task, only the first 5 seconds from both enrollment and test utterances are used. When evaluating performance on diarization task, utterances from a random 2-8 speakers are selected to simulate test data of duration more than 30 minutes each. In addition, the Fisher corpus \cite{DBLP:conf/lrec/CieriMW04} of conversational telephone speech is used to train an ASR network for bottleneck feature extraction. 

We also evaluate PRISM's performance on pre-training using a private dataset. The private dataset consists of 2,000 labeled speakers and 600,000 unlabeled utterances. The data are recorded by mobile devices with sampling rate 16k. The language is mandarin Chinese.

\subsection{Experimental Setup}

During the pre-training phase of PRISM, each training sample is comprised of random selection of utterances from random number of speakers. The number of speakers that makes up one sample ranges from 1 to 4 and the number of utterances from each speaker ranges from 1 to 3. Similar data simulation methods are utilized for diarization downstream task. For the purpose of specific scenarios, a small portion of overlapping speech can be simulated. For speaker verification task, at every iteration two utterances from each speaker are selected. Balanced positive and negative pairs are constructed.  PRISM takes a 40-dimensional filter bank features and 100-dimensional ASR bottleneck features as input. The bottleneck features are extracted from the ASR network trained on 40-dimensional high resolution MFCC features.

\section{Results \& Discussions}

Results from Table \ref{tab:NIST} suggest that PRISM outperforms most state-of-the-art speaker verification systems on short utterances. Table \ref{tab:private} indicates that PRISM has promising potential for utilizing large amounts of unlabeled data. 

\begin{table}[h]
    \caption{EER comparison on various systems, evaluated on NIST SRE 10, where enroll and test utterances are 5 seconds.}
    \centering
    \begin{tabular}{|c|c|}
    \hline
      System    & EER(\%) \\
      \hline
    x-vector  & 15.17 \\ \hline
    ETDNN  & 14.45 \\ \hline
    ResNet34  & 18.24 \\ \hline
    ECAPA-TDNN\cite{DBLP:conf/interspeech/DesplanquesTD20}  & 12.52 \\ \hline
    PRISM &\textbf{10.79} \\
    \hline
    \end{tabular}
    \label{tab:NIST}
\end{table}

\begin{table}[h]
    \caption{EER comparison on private mobile device dataset.}
    \centering
    \begin{tabular}{|c|c|}
    \hline
      System    & EER(\%) \\
      \hline
    x-vector  & 2.86 \\ \hline
    PRISM trained on only labeled data & 2.19 \\ \hline
    PRISM  with pre-training  &\textbf{1.39} \\
    \hline
    \end{tabular}
    \label{tab:private}
\end{table}

Table \ref{tab:clustering} describes the performance of several clustering methods. The first column, counting accuracy, measures the percentage of correct estimations of total number of speakers in a meeting. For k-means and spectral clustering, we use eigen-system to estimate the number of speakers\cite{DBLP:conf/nips/Zelnik-ManorP04}. For agglormerative hierarchical  clustering (AHC) the number of speakers is given by the resulting number of classes once the stopping criteria is met. Precision and recall are used to measure how accurate each system assigns segments to the corresponding clusters. 

\begin{table}[h]
    \caption{Clustering performance comparison on 1000 simulated meetings of 1 to 8 participants.}
    \centering
    \begin{tabular}{|c|c|c|c|}
    \hline
      Methods    & Cnt. Acc.(\%) & Prec. (\%)& Rec.(\%) \\
      \hline
    k-means  & 35.1 & 84.0 & 85.7 \\ \hline
    spectral  & 35.1 & 87.9 & 84.6 \\ \hline
    AHC  & 17.7 & 91.4 & 92.6\\ \hline
    PRISM-HDBSCAN &\textbf{92.9} & \textbf{96.7} & \textbf{98.1} \\
    \hline
    \end{tabular}
    \label{tab:clustering}
\end{table}

Both x-vector clustering system and end-to-end diarization system with encoder-decoder based attractors are compared to our proposed system, as listed in table \ref{tab:DER}. The EEND-EDA decreases significantly as the number of participants increases. The proposed system combines PRISM-based pairwise scoring and PRISM-E2E post-processing. It also incorporates the best-performing PRISM-HDBSCAN clustering method.  

\begin{table}[h]
    \caption{Comparison of three diarization systems, in terms of DER(\%). The evaluation set is simulated from NIST SRE10 dataset and each meeting lasts around 30 minutes.}
    \centering
    \begin{tabular}{|c|c|c|c|c|c|}
    \hline
     Methods & \multicolumn{5}{c|}{Number of speakers} \\ 
     \cline{2-6}
      & 4 & 5 & 6 & 7 & 8\\
      \hline
    x-vector \& k-means  & 10.1 & 13.9 & 17.2 & 20.3 & 25.2 \\ \hline
    EEND-EDA & 8.3 & 15.3 & 22.1 & 28.5 & 35.6 \\ \hline
    \textbf{Proposed}  &\textbf{7.9} & \textbf{9.8} & \textbf{13.7} & \textbf{16.1} & \textbf{18.4}\\
    \hline
    \end{tabular}
    \label{tab:DER}
\end{table}

The results from ablation studies are displayed in table \ref{tab:ablation}. If we directly concatenate phonetic information to input features in x-vector system, the improvement is marginal. On the other hand, if we train PRISM without using any phonetic information, we observe a $12$ percent relative reduction in EER, which is not as impressive as expected. However, if PRISM is trained with phonetic information concatenated to the input features, we observe a significant improvement. This result suggests that PRISM has managed to take advantage of phonetic information and regularize on the bias caused by content of speech. Fixed-embedding and softmax-based approaches, on the contrary, are not suitable for regularizing phonetic contents.

\begin{table}[h]
    \caption{Ablation studies on NIST SRE 10, where enroll and test utterances are both 5 seconds.}
    \centering
    \begin{tabular}{|c|c|}
    \hline
      System    & EER(\%) \\
      \hline
    x-vector  & 15.17 \\ \hline
    x-vector with phonetic information & 14.83 \\ \hline
    PRISM without phonetic information  & 13.44 \\ \hline
    PRISM with phonetic information &\textbf{10.79} \\
    \hline
    \end{tabular}
    \label{tab:ablation}
\end{table}

\section{Conclusion}

There are three major contributions of this work. First, we propose an indeterminate speaker representation model that benefits several downstream tasks such as speaker verification, speaker diarization, and speaker clustering. We also show that this approach can leverage large amounts of unlabeled data and introduces further improvements. Second, we propose a global-local diarization system based on PRISM. Third, we discovered that HDBSCAN outperforms other clustering methods such as k-means, spectral clustering, and AHC. Hence we made HDBSCAN the clustering backend for our PRISM pairwise similarities. 

For future works we are interested in investigating PRISM's effects on the other speech related tasks such as target speaker extraction, overlapping speech detection, etc. Even though the computation cost is less of a concern for 1-1 speaker verification task, it requires substantial reduction for clustering task. In addition, other representations of phonetic information are worthy of further exploration.

\bibliographystyle{IEEEtran}
\bibliography{refs,mybib}

\end{document}